\documentstyle[aps,eqsecnum]{revtex}
\begin{document}

\newcommand{\be}{\begin{equation}}
\newcommand{\ee}{\end{equation}}
\newcommand{\bea}{\begin{eqnarray}}
\newcommand{\eea}{\end{eqnarray}}
\newcommand{\nn}{\nonumber}

\title{Unconstrained Hamiltonian Formulation of  $SU(2)$ Gluodynamics}
\author{A. M. Khvedelidze \ $^a$
\thanks{Permanent address: Tbilisi Mathematical Institute,
380093, Tbilisi, Georgia} and
\,\, H.-P. Pavel\ $^b$ }
\address{$^a$ Bogoliubov Theoretical Laboratory, Joint Institute for Nuclear
Research, Dubna, Russia}
\address{$^b$ Fachbereich Physik der Universit\"at Rostock,
              D-18051 Rostock, Germany}
\date{January 27, 1999}
\maketitle

\begin{abstract}
$SU(2)$ Yang-Mills field theory is considered in the framework of the
generalized Hamiltonian approach and the equivalent unconstrained
system is obtained using the method of Hamiltonian reduction.
A canonical transformation to a set of adapted coordinates 
is performed in terms of which the Abelianization of the Gauss law constraints
reduces to an algebraic operation and the pure gauge degrees of freedom
drop out from the Hamiltonian after projection onto the constraint shell.
For the remaining gauge invariant fields two representations are
introduced where the three fields which transform as scalars under 
spatial rotations are separated from the three rotational fields.
An effective low energy nonlinear sigma model type Lagrangian is derived
which out of the six physical fields involves only one of the 
three scalar fields and two rotational fields summarized in a unit vector. 
Its possible relation to the effective Lagrangian proposed recently 
by Faddeev and Niemi is discussed. 
Finally the unconstrained analog of the well-known nonnormalizable 
groundstate wave functional which solves the Schr\"odinger equation with zero 
energy is given and analysed in the strong coupling limit.
\end{abstract}

\bigskip

\bigskip

\pacs{PACS numbers: 11.10.Ef, 11.15.-q, 11.15.Me}

\bigskip
\bigskip

\section{Introduction}

One of the main issues in the Hamiltonian formulation of Yang-Mills theories
is to find the projection from the phase space of canonical variables
constrained by the non-Abelian Gauss law
to the ``smaller''  phase space of unconstrained  gauge invariant
coordinates only.
Dealing with the problem of the elimination of the pure gauge
degrees of freedom two approaches exist,
the perturbative  and the nonperturbative one, 
with complementary features.
The conventional perturbative gauge fixing method
works successfully for the description of high energy phenomena,
but fails in applications in the infrared region.
The correct nonperturbative reduction of gauge theories
\cite{GoldJack}-\cite{KhvedZero}, on the other hand, leads to
representations for the unconstrained Yang-Mills systems,
which are valid also in the low energy region, but unfortunately up to now 
have been rather complicated for practical calculations.
The guideline of these investigations is the search for a
representation of the gauge invariant variables which is suitable 
for the description the infrared limit of Yang-Mills theory.
To get such a representation for the unconstrained system
we are following the  Dirac generalized Hamiltonian
formalism \cite{DiracL}-\cite{HenTeit}
using the method of Hamiltonian reduction  
(see \cite{Shanmugad}-\cite{GKP} and references therein) 
instead of the conventional gauge fixing
approach \cite{FadSlav}.
In previous work \cite{GKMP} it was demonstrated that for the case of the
mechanics of spatially constant $SU(2)$ Dirac-Yang-Mills
fields an unconstrained Hamiltonian can be derived which has a simple
practical form.
The elimination of the gauge degrees of freedom has been achieved
by performing a canonical transformation to a new adapted coordinates, 
in terms of which both the Abelianization of the Gauss law constraints
and the projection onto the constraint shell is simplified. The obtained
unconstrained system then describes the dynamics of a symmetric second
rank tensor under spatial rotations. The main-axis-transformation of this
symmetric tensor allowed us to separate the gauge invariant variables
into scalars under ordinary space rotations and into ``rotational'' degrees
of freedom. In this final form the physical Hamiltonian can be quantized
without operator ordering ambiguities.

In this work we shall generalize our approach from non-Abelian
Dirac-Yang-Mills mechanics \cite{GKMP} to field theory.
We shall give a Hamiltonian formulation of classical $SU(2)$ Yang-Mills
field theory entirely in terms of gauge invariant variables,
and separate these into scalars under ordinary space
rotations and into ``rotational'' degrees of freedom.
It will be shown that this naturally leads to their identification
as fields with ``nonrelativistic spin-2 and spin-0''.
Furthermore the separation into scalar and rotational degrees of freedom
will turn out to be very well suited for the study of the infrared limit
of unconstrained Yang-Mills theory.
We shall obtain an effective low energy theory involving only two of the 
three rotational fields and one of the tree scalar fields, and shall 
discuss its possible relation to the effective soliton Lagrangian proposed 
recently in \cite{FadNiem}.
Finally we shall analyse the well-known exact, but
nonnormalizable, solution \cite{Loos} of the functional Schr\"odinger
equation with zero energy in the framework of the unconstrained
formulation of the $SU(2)$ Yang-Mills theory.

The outline of the article is as follows.
In Section II we present the Hamiltonian reduction of $SU(2)$ Yang-Mills
field theory. We perform the canonical transformation to a new set of
adapted coordinates, Abelianize the Gauss law constraints, and achieve
the unconstrained description for $SU(2)$ Yang-Mills theory.
In section III two representations for the physical field in terms
scalars and rotational degrees are described.
Section IV is devoted to the study of the infrared limit of
unconstrained gluodynamics.
In Section V the well-known nonnormalizable solution of the functional
Schr\"odinger equation with zero energy is analysed in our unconstrained
formulation of the $SU(2)$ Yang-Mills theory.
Finally, in Section VI, we give our Conclusions.
In an Appendix we list several formulae
for nonrelativistic spin-0, spin-1 and spin-2 used in the text.

\section{Reduction of gauge degrees of freedom}

The degenerate character of the conventional Yang-Mills action
for \(SU(2)\)  gauge fields \( A^a_{\mu}(x) \)
\be \label{eq:act}
{\cal S} [A] : = - \frac{1}{4}\ \int d^4x\ F^a_{\mu\nu} F^{a\mu \nu}
\quad ,\quad \quad
F^a_{\mu\nu} : = \partial_\mu A_\nu^a  -  \partial_\nu A_\mu^a
+ g \epsilon^{abc} A_\mu^b A_\nu^c~,
\ee
requires the use of the generalized Hamiltonian approach \cite{DiracL}.
From the definition of the canonical momenta
$
P_{a}  :=  \partial L/\partial (\partial_0{A}^a_0 ) \,,\,\,\,\,\,\,\,
E_{ai} := \partial L/\partial (\partial_0 {A}_{ai})
$
it follows, that the phase space spanned by the variables
\( (A^a_0,  P^a) \),  \( (A_{ai}, E_{ai}) \)
is restricted by the three primary constraints
$P^a (x) = 0~$.
According to the Dirac procedure in this case the evolution of the system
is governed by the total Hamiltonian containing three arbitrary functions
\(\lambda_a (x)\)
\be \label{eq:totham}
H_T := \int d^3x\left[\frac{1}{2}\left( E_{ai}^2  + B_{ai}^2 (A) \right)
 - A^a_0  \left(\partial_i E_{ai} + g\epsilon_{abc}A_{bi}E_{ci}\right) +
\lambda_a (x) P^a (x)
\right]~,
\ee
where
$B_{ai}(A) :=
\epsilon_{ijk}\left(\partial_j A_{ak}+{1\over 2}g
\epsilon_{abc}A_{bj} A_{ck}\right)$
is  the non-Abelian magnetic field.
From the conservation  of the primary
constraints $P^a = 0~$ in time one obtains the non-Abelian Gauss
law constraints
\be  \label{eq:secconstr}
\Phi_a : = \partial_i E_{ai}+g\epsilon_{abc} A_{ci} E_{bi} = 0~.
\ee
Although the total Hamiltonian (\ref{eq:totham}) depends  on the
arbitrary functions $\lambda_a (x)$
it is possible to extract the dynamical variables
which have uniquely predictable dynamics. Furthermore they can be chosen
to be free of any constraints. Such an extracted system with predictable
dynamics without constraints is called unconstrained.

The non-Abelian character of the secondary constraints,
\bea  \label{eq:algb1}
\{\Phi_a (x) , \Phi_b (y)\} &= & g\epsilon_{abc}\Phi_c (x) \delta(x-y)~,
\eea
is the main obstacle for the corresponding projection to the
unconstrained phase space.
For Abelian constraints
\(\Psi_\alpha  \,\,  (\{\Psi_\alpha, \Psi_\beta\} = 0)\)
the projection to the reduced phase space
can be simply achieved in the following two steps. One performs a canonical
transformation to new variables such that part of the new momenta $P_\alpha$
coincide with the constraints $\Psi_\alpha $. After the projection onto the
constraint shell, i.e. putting in all expressions
$P_\alpha = 0$, the coordinates canonically conjugate to the $P_\alpha$
drop out from the physical quantities. The remaining
canonical pairs are then gauge invariant and form the basis for the
unconstrained system.
For the case of non-Abelian constraints (\ref{eq:algb1}) it is clearly
impossible to obtain such a canonical basis via canonical transformations
alone.
The way to avoid this difficulty is to replace the set of non-Abelian
constraints (\ref{eq:algb1}) by a new set of Abelian constraints
which describe the constraint surface in phase space. This
Abelianization procedure reduces the problem to the Abelian case.
There are several methods of Abelianization of constraints
(see  e.g. \cite{GKPabel}, \cite{GKP}   and   references therein).
In general the procedure can be reduced to the
solution of  linear differential equations  with partial derivatives
\cite{GKPabel}.

\subsection{Canonical transformation $ \& $ Abelianization
     of the Gauss law constraints }

The problem of Abelianization is considerably simplified when
studied in terms of coordinates adapted to the action of the gauge group.
The knowledge of the $SU(2)$ gauge transformations
\be  \label{eq:tran}
A_\mu \,\,\, \rightarrow \,\,\,  A_\mu^{\prime}   =
U^{-1}(x) \left( A_\mu  - {1\over g}\partial_\mu \right) U(x)~,
\ee
which leaves the Yang-Mills action (\ref{eq:act}) invariant,
directly promts us with the choice of adapted coordinates by using the
following point transformation to the new set of Lagrangian coordinates
$ q_j\ \ (j=1,2,3)$  and the six elements
$Q_{ik}= Q_{ki}\ \ (i,k=1,2,3)$ of the positive definite
symmetric $3\times 3$ matrix $Q$
\be
\label{eq:gpottr}
A_{ai} \left(q, Q \right) :=
O_{ak}\left(q\right) Q_{ki}
- {1\over 2g}\epsilon_{abc} \left( O\left(q\right)
\partial_i O^T\left(q\right)\right)_{bc}\,,
\ee
where \( O(q) \) is an orthogonal $3\times 3$ matrix
parametrised by the $q_i$.
\footnote{
In the strong coupling limit the representation (\ref{eq:gpottr}) reduces
to the so-called polar representation for arbitrary quadratic matrices
for which the decomposition can be proven to be well-defined and
unique (see for example \cite{Marcus}).
In the general case we have the additional second term which takes
into account the inhomogeneity of the gauge transformation
and (\ref{eq:gpottr}) has to be regarded as a set of partial
differential equations for the $q_i$ variables.
The uniqueness and regularity of the suggested transformation
(\ref{eq:gpottr}) depends on the boundary conditions
imposed. In the present work the uniqueness and regularity
of the change of coordinates is assumed as a reasonable
conjecture without search for the appropriate boundary conditions.
}
In the following we shall show that in terms of these variables the
non-Abelian Gauss law constraints (\ref{eq:secconstr})
only depend on the \( q_i \) and their conjugated momenta \( p_i\)
and after Abelianization become \( p_i = 0 \).
The unconstrained variables $ Q_{ik} $ and their conjugate \( P_{ik} \) are
gauge invariant, i.e. commute with the Gauss law, and represent the basic
variables for all observable quantities.\footnote{The freedom to use
other canonical variables
in the unconstrained phase space corresponds to another fixation
of the six variables $Q$ in the representation (\ref{eq:gpottr}).
This observation clarifies the connection
with the conventional gauge fixing method. We shall discuss this point
in forthcoming publications (see also ref. \cite{Muz}).}
The transformation (\ref{eq:gpottr}) induces a point canonical 
transformation linear in the new canonical momenta \( P_{ik} \) 
and \(p_i \). Using the corresponding  generating functional
depending on the old momenta and the new coordinates,
\be
F_3 \left[ E; q, Q \right] :=  \int d^3z \ E_{ai}(z)
A_{ai} \left(q(z), Q(z)\right)~,
\ee
one can obtain the  transformation to new canonical momenta
\( p_i \) and \( P_{ik} \)
\bea  \label{eq:mom1}
p_j (x)& :=  & \frac{\delta F_3 }{\delta q_j(x)}
= - {1\over g} \Omega_{jr}\left(D_i(Q)O^TE\right)_{ri}~,\\
\label{eq:mom2}
P_{ik}(x)& :=  &\frac{\delta F_3}{\delta Q_{ik}(x)}
= \frac{1}{2}\left(E^TO + O^T E \right)_{ik}~.
\eea
Here
\be
\label{Omega}
\Omega_{ji}(q) \, : = \,-\frac{i}{2}
\mbox{Tr}\ \left(
O^T\left(q\right)\frac{\partial O \left(q\right)}
{\partial q_j}
\, J_i \right),
\ee
with  the \( 3 \times 3 \) matrix generators of \(SO(3)\),
\( \left( J_i\right)_{mn} := i\epsilon_{min} \),
and the corresponding covariant derivative $D_i(Q)$
in the adjoint  representation
\be\label{eq:newcon}
\left(D_i(Q)\right)_{mn} := \delta_{mn}\ \partial_i
-ig \left( J^k \right)_{mn}\ Q_{ki}.
\ee
A straightforward calculation based on the  linear
relations  (\ref{eq:mom1}) and (\ref{eq:mom2})
between the old and the new momenta
leads to the following expression for the field strengths $E_{ai}$
in terms of the new canonical variables
\be  \label{eq:elpotn}
E_{ai} = O_{ak}\left(q\right) \biggl
[\,  P_{\ ki} +\epsilon _{kis}
 {}^\ast D^{-1}_{sl}(Q)
\left[
\left(\Omega^{-1} p \right)_{l} -
{\cal S}_l\,\right]\,\biggr]~.
\ee
Here  \( {}^\ast D^{-1}\) is the inverse of the matrix operator
\be
\label{DeltaQ}
{}^\ast D_{ik}(Q) : = -i\left(J^m D_m(Q)\right)_{ik}~,
\ee
and
\be
\label{eq:spin}
{\cal S}_k (x) := \epsilon_{klm}\left(P Q\right)_{lm} -
{1\over g} \partial_l P_{kl}\,.
\ee

Using the representations (\ref{eq:gpottr}) and (\ref{eq:elpotn})
one can easily convince oneself that the
variables \( Q\) and \( P \) make no contribution to the
Gauss law constraints (\ref{eq:secconstr})
\be
\Phi_a = O_{as}(q) \Omega^{-1}_{\ sj}(q)
p_j = 0~.
\label{eq:4.54}
\ee
Here and in (\ref{eq:elpotn}) we assume that the matrix
$ \Omega$ is invertible. The equivalent set of Abelian constraints is
\be
p_a   = 0~.
\ee
They are Abelian due to the canonical structure of the new variables.

\subsection{The Hamiltonian in terms of unconstrained fields}

After having rewritten the model in terms of the new canonical coordinates
and after the Abelianization of the Gauss law constraints,
the construction of the unconstrained Hamiltonian system
is straightforward. In all expressions we can simply put
\(p_a = 0\).
In particular, the Hamiltonian in terms of the unconstrained
canonical variables \(Q\) and \(P\)
can be represented by the sum of three terms
\be \label{eq:uncYME}
H[Q,P]=
\frac{1}{2} \int d^3{x}
\biggl[\,
\mbox{Tr}(P)^2 + \mbox{Tr}(B^2(Q))
\ + {1\over 2}{\vec E}^2(Q,P)
\biggr]~.
\ee
The first term is the conventional quadratic ``kinetic'' part and
the second the ``magnetic potential'' term
which is the trace of the square of the non-Abelian magnetic field
\be
 B_{sk}: = \epsilon_{klm}
\left( \partial_l Q_{sm} +
\frac{g}{2}\,  \epsilon_{sbc} \, Q_{bl}Q_{cm} \right)~.
\ee
It is intersting that after the elimination of the pure gauge degrees of
freedom the magnetic field strength tensor is the commutator of the
covariant derivatives (\ref{eq:newcon})
$F_{ij} = [D_i(Q), D_j(Q)]$.

\noindent
The third, nonlocal term in the Hamiltonian (\ref{eq:uncYME}) is the square
of the antisymmetric part of the electric field (\ref{eq:elpotn}),
$E_{s} :=(1/2)\epsilon_{sij}{E}_{ij}$,
after projection onto the constraint surface.
It is given as the solution of the system of differential
equations\footnote{
We remark that for the solution of this equation we need to impose 
boundary conditions only on the physical variables \(Q \), in 
contrast to Eq. (\ref{eq:gpottr}) for which boundary conditions only
for the unphysical variables $q_i$ are needed.}
\be
\label{vecE}
{}^\ast D_{ls}(Q) E_s = g{\cal S}_l~,
\ee
with the derivative ${}^\ast D_{ls}(Q)$ defined in (\ref{DeltaQ}).
Note that the vector ${\cal S}_i(x)$, defined in (\ref{eq:spin}),
coincides up to divergence terms with the spin
density part of the Noetherian angular momentum,
$ S_i (x) := \epsilon_{ijk}A_j^aE_{ak} $,
after transformation to the new
variables and projection onto the constraint shell.
\footnote{
Note that the presence of this divergence term destroys the
\( so(3)\) algebra of densities due to the presence of Schwinger terms
\be
\{{\cal S}_i (x),{\cal S}_j (y)\} = \epsilon_{ijk}{\cal S}_k (x)\delta(x-y)
+ \epsilon_{ijs}P_{sk} (x)\partial^x_k
\delta(x-y),
\ee
but maintains the value of spin and its algebra if one neglects
the surface terms.}
The solution ${\vec E}$ of the differential equation (\ref{vecE})
can be expanded in a $1/g$ series. The zeroth order term is
\be
\label{vecE1}
E^{(0)}_{s}=
\gamma^{-1}_{sk}\epsilon_{klm}\left(P Q\right)_{lm}~,
\ee
with $\gamma_{ik}:= Q_{ik}-\delta_{ik}\mbox{Tr}(Q)$,
and the first order term is determined as
\be
\label{vecE2}
E^{(1)}_{s} := {1\over g}
\gamma^{-1}_{sl}\left[(\mbox{rot}\ {\vec {E}}^{(0)})_l
-\partial_k P_{kl}\right]
\ee
from the zeroth order term.
The higher terms are then obtained by the simple recurrence relations
\be
\label{vecE3}
E^{(n+1)}_{s} := {1\over g}
\gamma^{-1}_{sl}(\mbox{rot}\ {\vec {E}}^{\ (n)})_s~.
\ee
One easily recognizes in these expressions
the  conventional definition of the covariant curl operation \cite{Martin}
in terms of the covariant derivative
$$
\mbox{curl}\,S(e_i, e_j) := \big< \nabla_{e_i} S,  e_j \big>  -
\big< \nabla_{e_j} S,  e_i \big>~,
$$
calculated in the basis $e_i : = \left (\gamma^{1/2}\right)_{ij}
{\partial}_j$  and $
\gamma_{ij}: = \big< e_i,  e_j \big>$ with the corresponding
 connection  $\nabla_{e_i}e_j  = \Gamma^l_{ij}e_l$, e.g.
\be
E^{(1)}_{ij} = \mbox{curl} \,S(e_i, e_j)~.
\ee

\section{The unconstrained Hamiltonian in terms of scalar and rotational
degrees of freedom}

In the previous section we have obtained the unconstrained Hamiltonian system
in terms of physical fields represented by a positive definite symmetric
matrix $Q$. The initial gauge fields $A_i$ transformed as vectors
under spatial rotations. We now would like to study the transformation
properties of the corresponding reduced matrix field $Q$.
For systems possessing some rigid symmetry
it is well known to be very useful for practical calculations
to pass to a coordinate basis such that a subset
of variables is invariant under the action of the symmetry group.
In this section we shall therefore carry out the explicit separation
of the rotational degrees of freedom, which vary under rotations,
from the scalars.

\subsection{Transformation  properties of the unconstrained fields under
space rotations}

In order to search for a parametrization of the unconstrained variables
in Yang-Mills theory adapted to the action of the group of spatial rotations
we shall study the corresponding transformation properties
of the field $Q$.
The total Noetherian angular momentum vector for $SU(2)$ gluodynamics is
\be
\label{eq:firstint}
I_i = \epsilon_{ijk} \int d^3 x~
\left( E_{aj}A_k^a +  x_k E_{al}\frac{\partial A_{al}}{\partial x^j}
\right)~.
\ee
After elimination of the gauge degrees of freedom it reduces to
\be
\label{IQP}
I_i = \int d^3 x~\epsilon_{ijk}
\left((P Q)_{jk}
+x_k \mbox{Tr}\left( P\partial_j Q \right)\right)~,
\ee
where surface terms have been neglected.

Under infinitisimal rotations in 3-dimensional space, $\delta x_i =
\omega_{ij}x_j$, generated by (\ref{IQP}), the physical field $Q$
transforms as
\be
\delta_{\omega} Q_{ij} =
\epsilon_{smn}\omega_{mn}\{ Q_{ij}, I_s \} =
\omega_{mn}\left(S^{mn} Q\right)_{ij}
+ {orbital~ ~part~~ transf.}
\ee
with the  matrices
\be
\left( S \right)_{(il)(sj)}^{mn}: =
\left( \delta_{il}\delta^m_k \delta^n_s
+ \delta^m_i \delta^n_l \delta_{sj}\right) -
(m \leftrightarrow n)~,
\ee
which describes the $SO(3)$ rotations of a 3-dimensional second rank tensor
field
\be
\label{Qtrafo}
Q^\prime_{ik}=R_{il}(\omega)R_{km}(\omega)Q_{lm}~.
\ee
It is well known that any symmetric second rank tensor can be decomposed
into its
irreducible components, one spin-0 and the five components of a spin-2 field
by extraction of its trace \cite{Brink}.
On the other hand it can be diagonalized via a main axis transformation,
which corresponds to a separation of diagonal fields, which are invariant
under rotations, from the rotational degrees of freedom.
In the following paragraphs we shall investigate both representations
and their relation to each other.

\subsection{The unconstrained Hamiltonian in terms of spin-2 and spin-0
fields}

As shown in the preceeding paragraph the six independent elements of
the matrix field  $Q$ can be represented as a mixture of fields with
nonrelativistic spin-2 and spin-0.
In order to put the theory into a more transparent form explicitly showing
its rotational invariance, it is useful to
perform a canonical transformation to the corresponding spin-2
and spin-0 fields as new variables.
To achieve this let us decompose the symmetric matrix $Q$ into
the irreducible representations of S0(3) group
\be
\label{QY}
Q_{ij}(x) = {1\over \sqrt{2}}{Y}_A(x)\  T^{A}_{ij}
+{1\over \sqrt{3}} \Phi(x)\ I_{ij}~,
\ee
with the field \(\Phi \) proportional to the trace of $Q$
as spin-0 field and the 5-dimensional spin-2 vector ${\bf Y}(x)$ with
components $Y_A$ labeled by its value of spin
along the $z$- axis, $ A= \pm 2, \pm 1, 0 $.
\footnote{Everywhere in the article 3-dimensional
vectors are topped by an arrow and their Cartesian and spherical
components are labeled by small Latin and Greek letters respectively,
while the 5-dimensional spin-2 vectors are written in boldface and their
``spherical'' components labeled by capital Greek letters.
For the lowering and raising of the indices of
5-dimensional vectors the  metric tensor
$\eta_{AB}= (-1)^A\delta_{A,-B}$ is used.}
$I$ is the $3\times 3$ unit matrix and the five traceless $3\times 3$
basis matrices ${\bf T}^A$ are listed in the Appendix.

The  momenta $P_A(x)$ and $P_\Phi(x)$ canonical conjugate to the
fields $Y_A(x)$ and $\Phi(x)$ are the components
of the corresponding expansion for the $P$ variable
\be
\label{PY}
P_{ij}(x) = {1\over \sqrt{2}} P_A(x)\ T^{A}_{ij}
+ {1\over \sqrt{3}} P_\Phi(x) \ I_{ij}~.
\ee
For the magnetic field $B$ we obtain the expansion
\be
B_{ij}(x) ={1\over \sqrt{2}} {H}_A(x)\ T^{A}_{ij} +
{1\over \sqrt{2}} h_\alpha(x)\ J^{\alpha}_{ij} + {1\over \sqrt{3}}b(x)\
I_{ij}~,
\ee
with the components
\bea \label{eq:magspin}
&& H_A :={1\over 2} c^{(2)}_{A \beta B} \partial_\beta{Y}^B
+ {g\over \sqrt{3}}\left(\frac{1}{\sqrt{2}} {}^\ast Y_A -
\Phi Y_A\right)~, \\
&& h_\alpha: = {1\over 2} d^{(1)}_{\alpha B\gamma} \partial_\gamma{Y}^B
  + \sqrt{2\over 3}\partial_\alpha\ \Phi~,\\
&& b : =  \frac{g}{\sqrt{3}}({1\over 2}{\bf Y}^2 - {\Phi}^2)~,
\eea
in terms of the structure constants $c^{(2)}_{A\beta C}$
and $d^{(1)}_{\alpha B \gamma }$ of the algebra of the spin-1  matrices 
$J^{\alpha}$
and the spin-2 matrices ${\bf T}^A$, listed in the Appendix,
and another five-dimensional vector
\be
{}^{\ast}Y_{C}:=  d^{(2)}_{{CAB}} {Y}^A{Y}^B~,
\ee
with constants $d^{(2)}_{ABC}$ given explicitly in the Appendix.
Finally we obtain  the reduced Hamiltonian in terms of
spin-2 and spin-0 field components
\be
\label{unchyphi}
H[{\bf P}, {\bf Y}, P_\Phi, \Phi]: =\frac{1}{2}
\int d^3x\ \left({\bf P}^2(x) + {\vec E}^2(x) + P^2_{{}_{\Phi}}(x) +
{\bf H}^2(x) +{\vec h}^2(x) +b^2(x)\right),
\ee
with expressions (\ref{eq:magspin}) for the magnetic field components and
the antisymmetric part ${\vec E}$ of the electric field given by
(\ref{vecE1}) - (\ref{vecE3}), expressing
$Q$ and $P$ in terms of ${\bf Y}$, $\Phi$ and ${\bf P}$, 
$P_{\Phi}$ via (\ref{QY}) and (\ref{PY}).
In order to discuss the transformation properties of the spin-2 fields
${\bf Y}$ under spatial rotations
we rewrite the angular momentum vector (\ref{IQP})
in terms of the fields ${\bf Y},{\bf P}$ and $\Phi,P_\Phi$
\be
I_i =  S_i
+\epsilon_{ijk} \int d^3{x}~ x_j (
P_\Phi\partial_k \Phi +
{P}_A\partial_k {Y}^A)~,
\ee
with the spin part
\be
S_i= i({\bf J}_i)_{AB} Y^A P^B~.
\ee
Here the three $5 \times 5$ matrices ${\bf J}^k$ are the elements
of the $so(3)$ algebra. They are shown explicitly in the Appendix.
The $I_i$ generate the transformation of the 5-dimensional vector $Y$
under infinitisimal  rotations in 3-dimensional space
$\delta x_i = \epsilon_{ijk}\omega_k x_j$
\be
\delta_{\omega} Y^{A} = \omega_k \{Y^A, S_k\} = -i ({\bf J}^k)_{AB} Y^B~.
\ee
For finite spatial rotations $R(\omega)$ we therefore have
\be
\label{Ytrafo}
Y'_A= D_{AB}(\omega) Y_B~,
\ee
with the well-known 5-dimensional spin-2 $D$-functions \cite{Brink}
related to the $3\times 3$ orthogonal matrix $R(\omega)$ via
\be
\label{DR}
D_{AB}(\omega) = \mbox{Tr}(R(\omega){\bf T}_AR^T(\omega){\bf T}_B)~.
\ee
The transformation rule (\ref{Ytrafo}) is in accordance with (\ref{Qtrafo}).

Note that for a complete investigation of the transformation properties
of the reduced matrix field $Q$ under the whole Poincar\'e group
one should also include the Lorentz transformations.
But we shall limit ourselves here to the isolation of the scalars under
spatial rotations and can
treat $Q$ in terms of ``nonrelativistic  spin-0 and spin-2
fields'' in accordance with the conclusions obtained in the work
\cite{Faddeev79}.
The study of the nonlinear representation of the whole Poincar\'e group
in terms of the unconstrained variables will be the subject of further
investigation.

\subsection{Separation of scalar and rotational degrees of freedom}

In this paragraph we would like to introduce a parametrization of the
5-dimensional ${\bf Y}$ field in terms of three Euler angles and two
variables which are invariant under spatial rotations.
The transformation property (\ref{Ytrafo}) prompts us with the
parametrization
\be
Y_A(x) = {D}_{AB}(\chi(x)) M^{B}(x)~,
\ee
in terms of the three Euler angles $\chi_i=(\phi,\theta,\psi)$
and some 5-vector ${\bf M}$. The special choice
\be
{ \bf M}(\rho,\alpha) =\rho \left( -{1\over \sqrt{2}}\sin\alpha,\ 0, \
\cos\alpha, \ 0,\ -{1\over \sqrt{2}}\sin\alpha\right)
\ee
corresponds to the main-axis-transformation of
the original symmetric $3\times 3$ matrix field $Q(x)$,
\be
\label{mainax}
Q (x)=
R^T(\chi(x))Q_{\rm diag}\left(\phi_1(x),\phi_2(x),\phi_3(x)\right)
R(\chi(x))~,
\ee
with the $D(\chi)$ related to $R(\chi)$ via (\ref{DR})
and the rotational invariant variables $\Phi,\rho,\alpha$ related to the
diagonal elements $\phi_i$ via
\footnote{Similar variables have been used as density and deformation
variables in the collective model of Bohr in Nuclear Physics \cite{BoMo} and
as a parametrization for the square of the eigenvalues of the rotational
invariant part of the gauge field by \cite{MarVau} in the representation
proposed in \cite{Simon}.
}
\bea
&&\phi_1 :=\frac{1}{\sqrt{3}}\Phi
+ \sqrt{2\over 3}\ \rho\cos(\alpha +\frac{2\pi}{3})~,\nonumber\\
&&\phi_2 := \frac{1}{\sqrt{3}}\Phi
+ \sqrt{2\over 3}\ \rho\cos(\alpha +\frac{4\pi}{3})~,\nonumber\\
&&\phi_3 :=  \frac{1}{\sqrt{3}}\Phi
+ \sqrt{2\over 3}\  \rho\cos\alpha~.
\label{phi123}
\eea
As it was mentioned in the first part of the paper, the matrix $Q$
is symmetric  positive definite.
The variables $\phi_i$ are therefore positive
\be
\label{phipos}
\phi_i \ge 0~,\ \ i=1,2,3~,
\ee
and the domain of definition for the variables $\alpha$ and $\rho$ can
correspondingly be taken as
\be
0\le \rho \le \sqrt{2}\Phi \ ,\ \ \ \ \ \ \ \alpha \le {\pi\over 3}~.
\ee
The main-axis-transformation of the symmetric second rank tensor field
$Q$ therefore induces a parametrization of the five spin-2 fields $Y^A$
in terms of the three rotational degrees of freedom, the Euler angles
$\chi_i=(\psi,\theta,\phi)$, which describe
the orientation of the ``intrinsic frame'', and the two invariants $\rho$ 
and $\alpha$ represented by the 5-vector ${\bf M}$.
As the three scalars under spatial rotations we can hence use either
$\rho$, $\alpha$, and the spin-0 field $\Phi$, or the three
fields $\phi_i\ (i=1,2,3)$.

In the following we shall use the main-axis-representation (\ref{mainax}).
The momenta $\pi_i$ and $p_{\chi_i}$, canonical conjugate
to the diagonal elements $\phi_i$ and the Euler angles
$\chi_i$, can
easily be found using the generating function
\be
F_3 \left[ \phi_i, \chi_i; \ P \right]  : =
\int d^3 x\  \mbox{Tr}\left(Q P \right) =
\int d^3 x \ \mbox{Tr}\
\left( {R }^T(\chi) Q_{\rm diag}(\phi) {R }(\chi) P \right)
\ee
as
\bea
&&
\pi_i(x) = \frac{\partial {F_3}}{\partial \phi_i(x)} =
 \ \mbox{Tr} \left( P  {R}^{T} \overline{\alpha}_{i} {R}
\right),\nn\\
&&
p_{\chi_i}(x) = \frac{\partial {F_3}}{\partial \chi_{i}(x)} =
 \mbox{Tr} \left(
\frac{\partial {R}^{T} }{\partial \chi_{i}}
R \, \left[ P Q - Q P\right]
\right).
\eea
Here  $\overline{\alpha}_{i}$ are the diagonal matricis
with the elements $(\overline{\alpha}_i)_{lm}=\delta_{li}\delta_{mi}$.
Together with the  off-diagonal matricies
$(\alpha_i)_{lm}=|\epsilon_{ilm}|$
they form an orthogonal basis for symmetric  matrices, shown
explicitly in the Appendix.
The original physical momenta  \( P_{ik} \) can then be expressed
in terms of the new canonical variables  as
\bea \label{eq:newmom}
P (x) =
{R }^T(x) \left(
         \sum_{s=1}^3  \pi_s(x) \, \overline{\alpha}_{s} +
 {1\over \sqrt{2}}\sum_{s=1}^3 {\cal P}_s(x) \,\alpha_{s}\right)
{R }(x)~,
\eea
with
\be
{\cal P}_{i}(x):=\frac{\xi_i(x)}{\phi_j(x) - \phi_k(x)}
 \,\,\,\,\,
(cyclic\,\,\,\, permutation \,\,\, i\not=j\not= k )~,
\ee
where the $\xi_i $ are given in terms of Euler angles
$\chi_i=(\psi,\theta,\phi)$ as
\be
\xi_k(x): ={\cal M}(\theta, \psi)_{kl} p_{\chi_l}~,
\ee
\be \label{eq:MCmatr}
{\cal M}(\theta, \psi): =
\left (
\begin{array}{ccc}
\sin\psi/\sin\theta,     & \cos\psi,     & - \sin\psi\cot\theta     \\
-\cos\psi/\sin\theta,    & \sin\psi ,    &  \cos\psi\cot\theta      \\
            0,                  &   0 ,         &     1
\end{array}
\right ).
\ee
Note that the $\xi_i $ are the $SO(3)$ left-invariant Killing vectors
satisfying the algebra 
\bea
\{\xi_i(x),\xi_j(y)\}=-\epsilon_{ijk}\xi_k(x)\delta(x-y)~.
\nonumber
\eea
The antisymmetric part ${\vec E}$ of the electric field appearing in the
unconstrained Hamiltonian (\ref{eq:uncYME})
is given by the following expansion in a $1/g$ series, analogous to
(\ref{vecE1}) - (\ref{vecE3}),
\be
E_{i} = R^T_{is}\sum_{n=0}^{\infty} {\cal E}^{(n)}_{s}~,
\ee
with the zeroth order term
\be
{\cal E}^{(0)}_{i} := -\frac{\xi_i}{\phi_j+ \phi_k}
 \,\,\,\,\,
(cycl.\,\,\,\, permut. \,\,\, i\not=j\not= k )~,
\ee
the first order term given from ${\cal E}^{(0)}$ via
\be
\label{calE1}
{\cal E}^{(1)}_{i} := {1\over g}\frac{1}{\phi_j+\phi_k}
\left[\left((\nabla_{X_j}\vec{{\cal E}}^{(0)})_k
-(\nabla_{X_k}\vec{{\cal E}}^{(0)})_j
\right)-\Xi_i\right]~,
\ee
with cyclic permutations of $ i\not=j\not= k $, and
the higher order terms of the expansion determined via the recurrence
relations
\be
{\cal E}^{(n+1)}_{i} := {1\over g}\frac{1}{\phi_j+\phi_k}
\left((\nabla_{X_j}\vec{{\cal E}}^{(n)})_k
-(\nabla_{X_k}\vec{{\cal E}}^{(n)})_j\right)~.
\ee
Here the components of the covariant derivatives $\nabla_{X_k}$
in the direction of the vector field $X_i(x):= R_{ik}\partial_k$,
 \be
(\nabla_{X_i}\vec{\cal E})_b :=
X_i {\cal E}_b + \Gamma^d_{{\ }ib}{\cal E}_d~,
\ee
are determined by the connection depending only on the Euler angles
\be
\label{3dimcon}
\Gamma^b_{{\ }i a} := \left( R X_i R^T\right)_{ab}~.
\ee
It is easy to check that the connection $\Gamma^b_{{\ }i a}$ can be
written in the form
\be
\Gamma^b_{{\ }i a} = i(J^s)_{ab} ({\cal M}^{-1})_{sk} X_i \chi_k~,
\ee
using the matrix ${\cal M}$ given
in terms of the Euler angles $\chi_i=(\psi,\theta,\phi)$
in (\ref{eq:MCmatr}),
which expresses the dual nature of the Killing vectors $\xi_i$
in (\ref{eq:MCmatr}) and the Maurer-Cartan one-forms $\omega^i$ defined by
\be
RdR^T =: \omega^iJ^i, \quad \quad \quad
\omega^i = ({\cal M}^{-1})^i_kd\chi_k~.
\ee
The source terms $\Xi_k$ in (\ref{calE1}), finally, are given as
\be
\Xi_1= \Gamma^1_{{\ }2 2}(\pi_1-\pi_2)+{1\over 2}X_1\pi_1
       -\Gamma^2_{{\ }2 3}{\cal P}_2-\Gamma^1_{{\ }2 3}{\cal P}_1
       -2\Gamma^1_{{\ }1 2}{\cal P}_3+X_2{\cal P}_3
       \ +\ (2 \leftrightarrow 3)~,
\ee
and its cyclic permutaions $\Xi_2$ and $\Xi_3$.

The unconstrained Hamiltonian therefore takes the form
\be \label{eq:unch2}
H \, = \,
\frac{1}{2} \int d^3x
\left( \sum_{i=1}^3\pi_i^2
 + {1\over 2}\sum_{cycl.}{\xi^2_i\over (\phi_j-\phi_k)^2} +
{1\over 2}{\vec{\cal E}}^{\ 2} + V  \right)~,
\ee
where the potential term $V$
\be
\label{Vinhom1}
V[\phi,\chi] = \sum_{i=1}^{3} V_i[\phi,\chi]
\ee
is the sum of
\be
\label{Vinhom2}
V_1[\phi,\chi]= \left(\Gamma^1_{{\ }12}(\phi_2-\phi_1)
                              -X_2 \phi_1\right)^2
+\left(\Gamma_{{\ }13}^1(\phi_3-\phi_1)
                      -X_3 \phi_1\right)^2
                   +\left(\Gamma_{{\ }23}^1\phi_3
                +\Gamma_{{\ }32}^1\phi_2-g\phi_2 \phi_3\right)^2~,
\ee
and its cyclic permutations .
We see that through the main-axis-transformation of the symmetric second rank
tensor field $Q$ the rotational degrees of freedom, the Euler angles
$\chi$ and their canonical conjugate momenta $p_\chi$, have been isolated
from the scalars under spatial rotations and appear in the unconstrained
Hamiltonian only via the three Killing vector fields $\xi_k$,
the connections $\Gamma$, and the derivative vectors $X_k$.

\section{The infrared limit of unconstrained $SU(2)$ gluodynamics}

\subsection{The strong coupling limit of the theory}


From the expression (\ref{eq:unch2}) for the  unconstrained  Hamiltonian
one can analyse the classical system in the strong coupling
limit up to order $O(1/g)$. Using the leading order (\ref{calE1}) of
the $\vec{{\cal E}}$ we obtain the Hamiltonian
\be
\label{eq:strongh}
H_{S} \, = \,
\frac{1}{2}\int d^3x\left(\sum_{i=1}^3 \pi_i^2
 + \sum_{cycl.}\xi^{2}_i\frac{\phi_j^2+\phi_k^2 }{(\phi_j^2-\phi_k^2)^2}
+ V[\phi,\chi] \right)~.
\ee
For spatially constant fields the integrand of this expression
reduces to the Hamiltonian of $SU(2)$ Yang-Mills mechanics considered
in previous work \cite{GKMP}.
For the further investigation of the low energy properties of $SU(2)$
field theory a thorough understanding of the properties of the 
leading order $g^2$ term in (\ref{Vinhom1}), containing no derivatives,
\be
\label{Vhomo}
V_{\rm hom}[\phi_i] = g^2[\phi_1^2\phi_2^2+\phi_2^2\phi_3^2
+\phi_3^2\phi_1^2]~,
\ee
is crucial. The stationary points of the
potential term (\ref{Vhomo}) are
\be
\phi_1=\phi_2=0\ ,\ \ \ \phi_3\ -\ {\rm arbitrary}~,
\ee
and its cyclic permutations.
Analysing the second order derivatives of the potential
at the stationary points one can conclude that
they form a continous line of degenerate absolute minima at zero energy.
In other words the potential has a ``valley'' of zero energy minima
along the line $\phi_1=\phi_2=0$.
They are the unconstrained analogs of the toron solutions \cite{Luescher}
representing constant Abelian field configurations with vanishing magnetic
field in the strong coupling limit.
The special point $\phi_1=\phi_2=\phi_3=0$ corresponds to the ordinary
perturbative minimum.

In terms of the variables $\rho$, $\Phi$ and $\alpha$ the homogeneous
potential (\ref{Vhomo}) reads
\be
\label{eq:potnd}
V_{\rm hom} := \frac{g^2}{3}\left(
\Phi^4 + {3\over 4}\rho^4 -\sqrt{2} \Phi\rho^3 \cos3\alpha \right)~,
\ee
showing that the $\alpha$ parametrizes the strength of the coupling
between the spin-0 and spin-2 fields.
The valley of minima
is given by $\rho = \sqrt{2}\Phi,\ \alpha=0,\ \Phi$ arbitrary,
and the perturbative vacuum  by
$\rho = \Phi=\alpha=0$.

For the investigation of configurations of higher energy it is necessary to
include the part of the kinetic term in (\ref{eq:strongh}) containing
the angular momentum variables $\xi_i$.
Since the singular points of this term just correspond to the
absolute minima of the potential there will a competition between an 
attractive and a repulsive force. At the balance point we shall have a 
local minimum corresponding to a classical configuration with higher energy.

\subsection{Nonlinear sigma model type effective action as the infrared limit
of the unconstrained system}

We would like to find in this paragraph the effective classical field theory
to which the unconstrained theory reduces in the limit of infinite
coupling constant $g$, if we assume that the classical system spontaneously
chooses one of the classical zero energy minima of the leading order $g^2$ 
part (\ref{Vhomo}) of the potential. 
As discussed in the proceeding section these classical minima include apart 
from the perturbative vacuum, where all 
fields vanish, also field configurations with one scalar field 
attaining arbitrary values. 
Let us therefore put without loss of generality 
(explicitly breaking the cyclic symmetry)
\be
\phi_1=\phi_2=0\ ,\ \ \ \phi_3\ -\ {\rm arbitrary}~,
\ee
such that the potential (\ref{Vhomo}) vanishes.
In this case the part of the potential (\ref{Vinhom1}) containing 
derivatives takes the form
\bea
V_{\rm inhom}=
&& \phi_3(x)^2\big[(\Gamma^2_{{\ }1 3}(x))^2+(\Gamma^2_{{\ }2 3}(x))^2
           +(\Gamma^2_{{\ }3 3}(x))^2
           +(\Gamma^3_{{\ }1 1}(x))^2+(\Gamma^3_{{\ }2 1}(x))^2
           +(\Gamma^3_{{\ }3 1}(x))^2 \big]\nonumber\\
&& +\big[(X_1\phi_3)^2+(X_2\phi_3)^2\big]
   +2\phi_3(x)\big[\Gamma_{{\ }3 1}^3(x) X_1\phi_3
                   +\Gamma_{{\ }3 2}^3(x) X_2\phi_3\big]~.
\eea
Introducing the unit vector
\be
n_i(\phi,\theta):=R_{3i}(\phi,\theta)~,
\ee
pointing along the 3-axis of the ``intrinsic frame'', one can write
\be
V_{\rm inhom}= \phi_3(x)^2 \left(\partial_i {\vec n}\right)^2
              +(\partial_i\phi_3)^2
 -(n_i \partial_i\phi_3)^2 - (n_i\partial_i n_j) \partial_j (\phi_3^2)~.
\ee
Concerning the contribution from the nonlocal term in this phase,
we obtain for the leading part of the electric fields
\be
{\cal E}^{(0)}_1= -\xi_1/\phi_3\ \ ,\ \ {\cal E}^{(0)}_2=-\xi_2/\phi_3~.
\ee
Since the third component ${\cal E}^{(0)}_3$ and ${\cal P}_3$ are singular
in the limit
$\phi_1,\phi_2\rightarrow 0$, it is necessary to have $\xi_3\rightarrow 0$.
The assumption of a definite value of $\xi_3$ is in accordance with the fact
that the potential is symmetric around the 3-axis for small $\phi_1$ and
$\phi_2$, such that the intrinsic angular momentum $\xi_3$
is conserved in the neighbourhood of this configuration.
Hence we obtain the following effective Hamiltonian up to  order $O(1/g)$
\be
H_{\rm eff} = {1\over 2}\int d^3x
\left[\pi_3^2+{1\over \phi_3^2}(\xi_1^2+\xi_2^2) +(\partial_i\phi_3)^2
+\phi_3^2(\partial_i{\vec n})^2
-(n_i \partial_i\phi_3)^2 - (n_i\partial_i n_j) \partial_j (\phi_3^2)\right]~.
\ee
After the inverse Lagrangian transformation we obtain the corresponding
nonlinear sigma model type effective Lagrangian for  the unit vector
${\vec n}(t,{\vec x})$ coupled to the scalar field $\phi_3(t,\vec{x})$
\be
\label{Leff}
L_{\rm eff}[\phi_3,{\vec n}]= {1\over 2}\int d^3x 
\left[(\partial_\mu \phi_3^2)^2+
       \phi_3^2(\partial_\mu {\vec n})^2
      +(n_i \partial_i\phi_3)^2 
      + n_i(\partial_i n_j) \partial_j (\phi_3^2)\right]~.
\ee
In the limit of infinite coupling the unconstrained field theory in terms of
six physical fields equivalent to the original $SU(2)$ Yang-Mills theory 
in terms of the gauge fields $A_\mu^a$ reduces therefore to an effective 
classical field theory involving only one of the three scalar fields and
two of the three rotational fields summarized in the unit vector $\vec{n}$.  
Note that this nonlinear sigma model type Lagrangian
admits singular hedgehog configurations of the unit vector field $\vec{n}$.
Due to the absence of a scale at the classical level, however, these are
unstable. 
Consider for example the case of one static monopole placed at the origin,
\be
n_i:= x_i/r~,\ \ \ \phi_3=\phi_3(r)~, \ \ \ r:=\sqrt{x_1^2+x_2^2+x_3^2}~.
\ee
Minimizing its total energy $E$ 
\be
E[\phi_3]=4\pi \int dr \phi_3^2(r)
\ee 
with respect to $\phi_3(r)$ we find the classical solution $\phi_3(r)\equiv 0$.
There is no scale in the classical theory.
Only in a quantum investigation a mass scale such as a nonvanishing value for
the condensate $<0|\hat{\phi}_3^2|0>$ may appear, which might be related to 
the string tension of flux tubes directed along the unit-vector field 
${\vec n}(t,{\vec x})$. 
The singular hedgehog configurations of such string-like directed flux tubes 
might then be associated with the glueballs.
The pure quantum object $<0|\hat{\phi}_3^2|0>$ might be realized as a squeezed 
gluon condensate \cite{BlaPa}.
Note that for the case of a spatially constant condensate,
\be
<0|\hat{\phi}_3^2|0>=:2 m^2= const.~,
\ee 
the quantum effective action corresponding to (\ref{Leff}) should 
reduce to the lowest order term of the effective soliton Lagangian discussed
very recently by Faddeev and Niemi \cite{FadNiem}
\be
\label{FN}
L_{\rm eff}[{\vec n}]= m^2\int d^3x (\partial_\mu {\vec n})^2~.
\ee
As discussed in \cite{FadNiem}, for the stability of these knots 
furthermore a higher order Skyrmion-like term in the derivative expansion 
of the unit-vector field ${\vec n}(t,{\vec x})$ is necessary.
To obtain it from the corresponding higher order terms in the strong coupling 
expansion of the unconstrained Hamiltonian (\ref{eq:unch2}) is under present 
investigation.
 
First steps towards a quantum treatment of the unconstrained formulation
obtained in the preceeding paragraphs will be undertaken in the next section.

\section{Quantum ground state wave functional
and the classical configuration of lowest energy}

In this section we shall discuss several interesting
relations between  the quantum ground state wave functional
for the unconstrained Hamiltonian system
and the corresponding classical state configurations with zero energy.

\subsection{Exact ground state solution of the Schr\"odinger equation}

For the original constrained system of $SU(2)$ gluodynamics in terms
of the gauge fields $A_i^a(x)$ with the Hamiltonian
\be
{\cal H}(A) := {1\over 2}\int d^3x\ \left( - \left(
\frac{\delta}{\delta A^a_{i}(x)}\right)^2
+ B^2(x)\right)
\ee
and the Gauss law operators
\be
{\cal G}^a(x) := \left(\partial_i\delta^a_b 
- g\epsilon^{abc}A_i^c(x) \right)
\frac{\delta}{\delta A^b_{i}(x)}
\ee
in the Schr\"odinger functional formalism, a physical state has to
satisfy both the functional Schr\"odinger equation and the Gauss law
constraints
\bea
\label{eq:SchrG}
&&{\cal H}\Psi[A] = E \Psi[A]~,\\
&&{\cal G}^a(x)\Psi[A]=0~.
\eea
Remarkably, an exact solution for the ground state wave functional
$\Psi[A]$ can be given \cite{Loos}
\be
\label{PsiA}
\Psi[A] = \exp{\left(- 8\pi^2 W[A]\right)}
\ee
in terms of the so called ``winding number functional''$W[A]$
defined as the integral over
3-space
\be
W[A]: = \int d^3x\  K_0(x)
\ee
of the zero component of the Chern-Simons secondary characteristic
class vector \cite{Deser}
\footnote{
Note that whereas the topological invariant, the Pontryagin index
\be
q[A]: = \int d^4x \partial_\mu K^\mu (A)~,
\ee
and the corresponding Pontryagin density
$\mbox{Tr}\left( {}^\ast F^{\mu\nu} F_{\nu\mu}\right)$,
are gauge invariant quantities, the Chern-Simons vector $K^\mu$ is not
gauge invariant.
}
\be
\label{Kmu}
K^\mu(A): = -\frac{1}{16\pi^2} \epsilon^{\mu\nu\sigma\kappa}
\mbox{Tr}\left(
F_{\nu \sigma} A_\kappa -\frac{2}{3}gA_\nu A_\sigma A_\kappa
\right)~.
\ee
Since $W[A]$ obeys the functional differential equation
\be
\frac{\delta}{\delta A^a_{i}(x)}W[A]= B^a_i(x)~.
\ee
the wave functional (\ref{PsiA}) satisfies the above Schr\"odinger equation.
However this exact solution for the functional Schr\"odinger
equation with the zero energy is known to be nonnormalizable
and hence does not seem to have a physical meaning \cite{Jackiw}.

In the following we shall now analyse, how such an exact solution arises
in our unconstrained formalism. 

\subsection{Exact ground state solution for the unconstrained Hamiltonian}

Quantizing the variables $Q$ and $P$ of the unconstrained
Hamiltonian (\ref{eq:uncYME}) analogously to the $A_i^a$ above
\footnote{Note that due to the positive
definiteness of the elements of the matrix field $Q$ we
have to solve the Schr\"odinger equation in a restricted domain of
functional space. Special boundary conditions have to be imposed on the
wavefunctional such that all operators are well-defined
(e.g. Hermicity of the Hamiltonian).
A discussion on this subject in gauge theories can be found e.g. in the
review \cite{ProkhShab}.}
we have
\be
{\cal H} ={1\over 2} \int d^3x\ \left( - \left(
\frac{\delta}{\delta Q_{ij}(x)}\right)^2
+ B^2(x) + {1\over 2}\vec{E}^2(Q,\frac{\delta}{\delta Q} )
\right)~,
\ee
and hence the functional Schr\"odinger equation
\be \label{eq:Schr}
{\cal H}\Psi[Q] = E \Psi[Q]~.
\ee
The Gauss law has already been implemented by the reduction
to the physical variables.

A corresponding exact zero energy solution can indeed be found 
for our reduced Schr\"odinger equation (\ref{eq:Schr}). 
For this we note the following two
important properties of the potential terms
present in the Schr\"odinger equation (\ref{eq:Schr}).
Firstly, the reduced magnetic field $B_{ij}(Q)$ can be written
as the functional derivative of the functional ${\cal W}[Q]$
\be
\label{Wast}
{\cal W}[Q]:={1\over 32\pi^2} \int d^3x
\left[\mbox{Tr}(BQ) - \frac{1}{12}g
\left(
(\mbox{Tr}(Q^3) +\mbox{Tr}^3(Q) -
2 \mbox{Tr}(Q)\mbox{Tr}(Q^{2})
\right)\right]
\ee
 as
\be
\frac{\delta}{\delta Q_{ij}(x)}{\cal W}[Q]=B_{ij}(x)~.
\ee
Furthermore, the nonlocal term in the Schr\"odinger equation (\ref{eq:Schr})
annihilates ${\cal W}[Q]$
\be
{\vec E}^2[Q, \frac{\delta}{\delta Q_{ij}(x)} ]
{\cal W}[Q] = 0~.
\ee
The last equation can easily be found to hold if one takes into account that
the magnetic field $B_i = {}^\ast F_{0i}$ satisfies the Bianchi identity
$D_{i}{}^\ast F_{0i} =0$.

Thus the corresponding ground state wave functional solution for the
unconstrained Hamiltonian is
\be
\label{PsiQ}
\Psi[Q] = \exp{\left(- 8\pi^2{\cal W}[Q]\right)}~.
\ee

In order to investigate the relation of our ${\cal W}[Q]$ to the
above winding number functional $W[A]$ we write the zero component
of the the Chern-Simons secondary characteristic class vector $K^\mu$,
given in (\ref{Kmu}), in terms of the new variables $Q$ and
$q_i$
\be
\label{K_0}
K^0 (Q,q ) =
 {\cal K}^0(Q)  - {1\over 24\pi^2}\epsilon^{ijk}\left[\frac{2}{3}g
 \mbox{Tr}\left(\Omega_i\Omega_j\Omega_k\right)
-\partial_i\mbox{Tr}\left( Q_j\Omega_k\right)\right]~.
\ee
The first term
\be
{\cal K}^0(Q) :=  -\frac{1}{16\pi^2}
\epsilon^{ijk} \mbox{Tr}\left(
F_{i j} Q_k -\frac{2}{3}g Q_i
Q_j Q_k \right)
\ee
is a functional only of the physical $Q$ of a form similiar to that
of the original Chern-Simons secondary characteristic class vector.
Here we have introduced the $SU(2)$ matrices
$Q_l := Q_{li}\tau_i$, with the Pauli matrices $\tau_i$, and
\be
\Omega_i(q) :=
{1\over g} U^{-1}(q)\partial_iU(q)={1\over g}
\Omega_{ls}(q)\tau^s \left({\partial q_l\over 
\partial x_i}\right)~,
\ee
with the $SU(2)$ matrices $U(q)$ related to the $3\times 3$
orthogonal matrix
$O(q)$ via $O_{ab}(q)=
{1\over 2}Tr(O(q)\tau_a
O^T(q)\tau_b)$
and the $3\times 3$ matrix $\Omega_{ij}$ defined in (\ref{Omega}).

We observe that the space integral over the first term coincides with
the above functional ${\cal W}[Q]$ of (\ref{Wast})
\be
\label{K_0int}
\int d^3x {\cal K}^0(Q) = {\cal W}[Q]~.
\ee

Using the usual boundary condition
\footnote{Note that for the behaviour of the unphysical variables
$q_i$ we have no information. For example the
requirement of the finiteness of the action usually used to fix the
behaviour of the physical fields does not apply for the unphysical
field $q_i$.
}
\be
\label{bc}
U(q)\longrightarrow \pm I\ ,
\ee
we see that the space integral over the second term is proportional
to the natural number n representing the winding of the mapping of
compactified three space into $SU(2)$
\be
{g^3\over 24\pi^2}\int d^3x\ \epsilon^{ijk}
\mbox{Tr}(\Omega_i\Omega_j\Omega_k)
= n~.
\ee
Assuming here the vanishing of the physical field $Q$ at spatial
infinity there is no contribution from the third term.
Hence we obtain the relation
\be
\Psi[A]=\exp[-{8\pi^2\over g^2}n]\Psi[Q]
\ee
between the groundstate wave functional (\ref{PsiA}) of the
extended quantization scheme and the reduced (\ref{PsiQ}).
We find that the winding number of the original gauge field $A$
only appears as an unphysical normalization prefactor originating
from the second term in
(\ref{K_0}), which depends only on the unphysical $q_i$.
Furthermore, we note that the power ${8\pi^2\over g^2}n$ is the classical
Euclidean action of $SU(2)$ Yang-Mills theory of
self-dual fields \cite{BPST} with winding number $n$.

The physical part of the wavefunction, $\Psi[Q]$, on the other hand
however, has the same unpleasant property as (\ref{PsiA}) that it is
nonnormalizable.
In order to shed some light on the reason for its nonnormalizability,
it is useful to limit to the homogeneous case
and to analyse the properties of $\Psi[Q]$ in the neighbourhood of
the classical minima of the potential.

\subsection{Analysis of the exact groundstate wave functional 
in the strong coupling limit}

In the strong coupling limit the groundstate wave functional (\ref{PsiQ})
reduces to the very simple form
\be
\label{Psihomo}
\Psi[\phi_1,\phi_2,\phi_3]=\exp\left[-g\phi_1\phi_2\phi_3\right]~.
\ee
This wave functional is obviously nonnormalizable.
In difference to the Abelian case, however, where an analogous nonnormalizable
exact zero energy solution exists  
\footnote{
In (Abelian) electrodynamics the unconstrained form of the corresponding
exact zero-energy groundstate wave functional is an exponential of
$\int dk k a_1(k) a_2(k)$, where $a_1, a_2$ are the (momentum space)
polarization modes of $A$. This false ground state is nonnormalizable 
due to the sign indefiniteness of the exponent. 
We thank the Referee for pointing this out to us.},
the exponent of (\ref{Psihomo}) is free of any sign ambiguities 
due to the positivity
of the symmetric matrix $Q$ in the polar representation (\ref{eq:gpottr}),
and hence the positivity of the diagonal fields $\phi_i$, see (\ref{phipos}).
In order to investigate the reason for the nonnormalizability of 
(\ref{Psihomo}) we analyze it near the classical zero-energy minima,
that is, without loss of generality, in the neighbourhood of the line
$\phi_1=\phi_2=0$ of minima of the classical potential (\ref{Vhomo}).
It is useful to pass from the variables $\phi_1$ and $\phi_2$ transverse to
the valley to the new variables $\phi_\perp$ and $\gamma$ via
\be
\phi_1=\phi_\perp \cos\gamma\ \ \ \ \ \ \phi_2=\phi_\perp\sin\gamma
\ \ \ \ \left( \phi_\perp \ge 0~,\ \ \ \ \ 0\le\gamma \le {\pi\over 2}\right)~.
\ee
The classical potential then reads
\be
V(\phi_3,\phi_\perp,\gamma)=g^2\left(\phi_3^2\phi_\perp^2+{1\over 4}
\phi_\perp^4\sin^2(2\gamma)\right)~,
\ee
and the groundstate wave function (\ref{Psihomo}) becomes
\be
\Phi[\phi_3,\phi_\perp,\gamma]=
\exp\left[-{g\over 2}\phi_3\phi_\perp^2\sin(2\gamma)\right]~.
\ee
We see that close to the bottom of the valley, for small $\phi_\perp$, the
potential is that of a harmonic oscillator
and the wave functional correspondingly a Gaussian with a maximum at the
classical minimum line $\phi_\perp=0$. The height of the maximum is constant
along the valley.
The nonnormalizability of the groundstate wave function in the infrared 
region is therefore due to the outflow of the
wave function with constant values along the valley to arbitrarily
large values of the field $\phi_3$. This may result in the formation of
condensates with macroscopically large fluctuations of the field amplitude.
To establish the connection between this phenomenon and the
model of the squeezed gluon condensate \cite{BlaPa} will be an
interesting task for further investigation.

\section{Concluding remarks}

Following the Dirac formalism for constrained Hamiltonian systems
we have formulated several representations
for the classical $SU(2)$ Yang-Mills gauge theory entirely in terms of
unconstrained gauge invariant local fields.
All transformations which have been used,
canonical transformations and the Abelianization of the constraints,
maintain the canonical structures of the generalized Hamiltonian
dynamics.
We identify the unconstrained field with a symmetric positive
definite second rank tensor field under spatial rotations.
Its decomposition
into irredusible representations under spatial rotations leads
to the introduction of two fields, a five-dimensional
vector field ${\bf Y}(x)$  and a scalar field $\Phi (x)$.
Their dynamics is governed by an explicitly rotational invariant
non-local Hamiltonian.
It is different from the local Hamiltonian obtained by Goldstone and Jackiw
\cite{GoldJack} as well as by
Izergin et.al. \cite{Faddeev79}. They used the so-called electric field
representation with vanishing antisymmetric part of the electric field.
A representation for the Hamiltonian with a nonlocal interaction
of the unconstrained variables similar to ours has been derived in the
work of Y. Simonov \cite{Simon} based on another separation of scalar
and rotational degrees of freedom.
Our separation of the unconstrained fields into scalars under spatial
rotations and into rotational degrees of freedom, however, leads to a
simpler form of the Hamiltonian, which in particular is free of operator
ordering ambiguities in the strong coupling limit.
Our unconstrained representation of the Hamiltonian furthermore
allows us to derive an effective low energy Lagrangian for the rotational 
degrees of freedom coupled to one of the scalar fields
suggested by the form of the classical potential
in the strong coupling limit. The dynamics
of the rotational variables in this limit is summarized
by the unit vector describing the orientation of the intrinsic
frame. Due to the absence of a scale in the classical theory the
singular hedgehog configurations of the unit vector field is found to be 
unstable classically. In order to obtain a nonvanishing value for the 
vacuum expectation value for one of the three scalar field operators, 
which would set a scale, a quantum treatment at least to one loop order is 
necessary and is under present investigation.  
For the case of a spatially constant scalar quantum condensate
we expect to obtain the first term of a derivative expansion proposed recently 
by Faddeev and Niemi \cite{FadNiem}. As shown in their work
such a soliton Lagragian allows for stable massive knotlike configurations
which might be related to glueballs. For the stability of the knots
higher order terms in the derivative expansion, such as the Skyrme type
fourth order term in \cite{FadNiem}, are necessary. Their derivation
in the framework of the unconstrained theory, proposed in this paper, is
under investigation.
First steps towards a complete quantum description have been done
in this paper. In particular, we have investigated the
famous groundstate wave functional,
which solves the Schr\"odinger  equation  with zero energy eigenvalue.
In conclusion we would like to emphasize that our
investigation of low energy aspects of non-Abelian gauge theories
directly in terms of the physical unconstrained fields
offers an alternative to the variational calculations based on the
gauge projection method \cite{Kogan,Diakonov}.

The reason for trying to construct the physical variables
entirely in internal terms without the use of any gauge fixing
is the aspiration to maintain all local and global properties of the initial
gauge theory.
Several questions in connection with the global aspects of the reduction
procedure are arising at this point.
In the paper we describe how to
project SU(2) Yang-Mills theory onto the constraint
shell defined by the Gauss law.
It is well known that the exponentiation of infinitisimal
transformations generated  by the  Gauss law operator
can lead only to homotopically trivial gauge transformations,
continiously deformable to unity.
However, the initial classical action is
invariant under all gauge transformations,
including the homotopically nontrivial ones.
What trace does the
existence of large gauge transformations leave on the unconstrained system?
First steps towards a clarification of these important issues have been
undertaken in this paper, a more complete analysis is under present
investigation.

\acknowledgments

We are grateful for discussions with D. Blaschke, S.A. Gogilidze, I.I. Kogan,
A. Kovner, M. Lavelle, D.M. Mladenov, V.P. Pavlov, V.N. Pervushin, 
G. R\"opke, P. Schuck, A.N. Tavkhelidze and J. Wambach.
Furthermore we thank the Referee for his constructive comments.
One of us (A.K.) would like to thank Professor G. R\"opke
for his kind hospitality in the group ''Theoretische Vielteilchenphysik''
at the Fachbereich Physik of Rostock University, where part of this work has
been done and acknowledges the Deutsche Forschungsgemeinschaft for providing
a stipend for the visit.
His work was also supported  by the Russian Foundation for
Basic Research under grant No. 96-01-01223.
H.-P. P. acknowledges support by the Deutsche Forschungsgemeinschaft
under grant No. Ro 905/11-2 and by the Heisenberg-Landau Program for
providing a grant for a visit at the Bogoliubov Theoretical Laboratory 
in the JINR Dubna.

\newpage
\appendix

\section{Notations and some formulas}

\subsection{Spin 1 matices and eigenvectors }

For generators of spin-1  obeying the  algebra
\( [ J_i, J_j] = i \epsilon_{ijk} \ J_k\) we use the following matrix
realizations
\[
{J}^{1} = i\left (
\begin{array}{ccc}
0      &   0      &     0      \\
0      &   0      &   - 1      \\
0      &   1      &     0
\end{array}
\right ),
\quad \quad
{J}^{2} = i\left (
\begin{array}{ccc}
0      &   0      &    1     \\
0      &   0      &    0     \\
-1     &   0      &    0
\end{array}
\right ),
\quad \quad
{J}^{3} = i\left (
\begin{array}{ccc}
0      &  - 1      &    0     \\
1      &    0      &    0     \\
0      &    0      &    0
\end{array}
\right )~.
\]

Furthermore the representation of rotations R($\chi$) in terms of 
Euler angles $\chi=(\theta,\psi,\phi)$ is used
\be
R(\psi,\theta,\phi)= e^{-i\psi J_3} e^{-i\theta J_1} e^{-i\phi J_3}~.
\ee
The eigenfunctions of $J^2$
and $J_3$ are
\[
{\vec e}_{+1} = \frac{1}{\sqrt{2}}\left(
\begin{array}{c}
-1         \\
-i     \\
0
\end{array}
\right ),
\quad \quad
{\vec e}_{0} = \left(
\begin{array}{c}
0         \\
0      \\
1
\end{array}
\right),
\quad \quad
{\vec e}_{-1} = \frac{1}{\sqrt{2}}\left(
\begin{array}{c}
1         \\
-i      \\
0
\end{array}
\right)~.
\]
which are orthogonal with respect to the  metric $\eta_{\alpha \beta}:=
(-1)^\alpha\delta_{\alpha, -\beta}$
\be
\left({\vec e}_{\alpha}\cdot{\vec e}_{\beta}\right) = \eta_{\alpha \beta}
\ee
and satisfy the completeness condition
\be
{e}_{\alpha}^i{e}_{\beta}^j\eta^{\alpha \beta} =\delta^{ij}~.
\ee

\subsection{Spin-0, spin-1 and spin-2 tensors basis}

To obtain a matrix representation for spin-0, spin-1 and spin-2 basis 
matrices we use the Clebsh-Gordon decomposition for the  direct product of
spin-1 eigenvectors  $e_i^\alpha$
into the irreducible components  $3\otimes3 = 0 \oplus 1 \oplus 2$.
To distinguish the matrices correspondiong to the different
spins we use boldface notation for spin 2.

For spin-0 they read explicitly

\[
 I_{0}: = \frac{1}{\sqrt{3}}\left(
{\vec e}_{0}\otimes {\vec e}_{0}
-{\vec e}_{1}\otimes {\vec e}_{-1}
-{\vec e}_{-1}\ \otimes {\vec e}_{1} \right)=
\frac{1}{\sqrt{3}}
 \left (
\begin{array}{ccc}
1      &    0       &   0     \\
0       &    1       &   0     \\
0      &    0       &   1
\end{array}
\right )~,
\]

for spin-1

\[
 J^{+}: = \left(
{\vec e}_{0}\ \otimes {\vec e}_{+1}-
{\vec e}_{+1}\otimes {\vec e}_{0} \right)=
\frac{1}{\sqrt{2}}
 \left (
\begin{array}{ccc}
0       &   0       &    1     \\
0       &   0       &    i     \\
-1      &   -i      &    0
\end{array}
\right )~,
\]
\[
 J^{-}: = \left(
{\vec e}_{-1}\ \otimes {\vec e}_{0}-
{\vec e}_{0}\otimes {\vec e}_{-1}  \right)=
\frac{1}{\sqrt{2}}
 \left (
\begin{array}{ccc}
0      &   0      &    1     \\
0      &   0       &   -i     \\
- 1    &   i      &    0
\end{array}
\right )~,
\]

\[
 J^{0}: = \left(
{\vec e}_{-1}\ \otimes {\vec e}_{1}-
{\vec e}_{1}\otimes {\vec e}_{-1} \right)=
 \left (
\begin{array}{ccc}
0      &   -i       &   0     \\
i      &    0       &   0     \\
0      &    0       &   0
\end{array}
\right )~,
\]

for spin-2
\[
{\bf T}_{+2} =
\sqrt{2}\left({\vec e}_{+1}\otimes {\vec e}_{+1}\right) = \frac{1}{\sqrt{2}}
\left (
\begin{array}{ccc}
1      &    i      &    0     \\
i      &   -1      &    0     \\
0      &    0      &    0
\end{array}
\right )~,
\quad \quad
{\bf T}_{-2} =
\sqrt{2}\left({\vec e}_{-1}\otimes {\vec e}_{-1}\right) = \frac{1}{\sqrt{2}}
\left (
\begin{array}{ccc}
1       &   -i      &    0     \\
-i      &   -1      &    0     \\
0       &   0       &    0
\end{array}
\right )~,
\]

\[
{\bf T}_{+1}: = \left(
{\vec e}_{+1}\otimes {\vec e}_{0}
+
{\vec e}_{0}\otimes {\vec e}_{+1} \right)=
\frac{1}{\sqrt{2}}
 \left (
\begin{array}{ccc}
0      &   0      &   -1  \\
0      &   0      &   -i   \\
-1     &   -i     &    0
\end{array}
\right )~,
\]

\[
{\bf T}_{-1}: = \left(
{\vec e}_{-1}\otimes {\vec e}_{0}
+
{\vec e}_{0}\ \otimes {\vec e}_{-1} \right)=
\frac{1}{\sqrt{2}}
 \left (
\begin{array}{ccc}
0      &   0      &    1     \\
0      &   0      &   -i     \\
1      &   -i     &    0
\end{array}
\right )~,
\]

\[
{\bf T}_{0}: = \frac{1}{\sqrt{3}}\left(
{\vec e}_{+1}\otimes {\vec e}_{-1}
+2{\vec e}_{0}\otimes {\vec e}_{0}
+{\vec e}_{-1}\ \otimes {\vec e}_{+1} \right)=
\frac{1}{\sqrt{3}}
 \left (
\begin{array}{ccc}
-1      &    0       &   0     \\
0      &   - 1       &   0     \\
0      &     0       &   2
\end{array}
\right )~.
\]
They obey the following orthonormality relations:
\be
\mbox{Tr}\ ({\bf T}_{A}{\bf T}_{B}) = 2\eta_{AB}~,
\quad \quad
\mbox{Tr}\ ({\bf T}_{A}{J}_{\alpha}) = 0~,
\quad \quad
\mbox{Tr}\ ({J}_{\alpha}{J}_{\beta}) = 2\eta_{\alpha\beta}~,
\ee
the completeness condition
\be
\frac{1}{10}\sum_{A}({\bf T}_{A})_{il}({\bf T}_{A})_{km} +
 (I_{0})_{il}(I_{0})_{km} = \frac{1}{4}\left(
\delta_{im}\delta_{lk} +\delta_{il}\delta_{mk}
\right)~,
\ee
and the following commutation and anticommutation  relation
\bea
&&[{\bf T}_{A},{\bf T}_{B}]_{+} =
\frac{4}{\sqrt{3}}\eta_{AB} I_0
+ {2\over \sqrt{3}}{d}^{(2)}_{ABC}{\bf T}^C, \\
&&[{\bf T}_{A},{\bf T}_{B}]_{-} = {c}^{(2)}_{AB\gamma}J^{\gamma}~;
\eea
\bea
&&[{J}_{\alpha},{J}_{\beta}]_{+} =
\frac{4}{\sqrt{3}}\eta_{\alpha \beta} I_0
+ {d}^{(1)}_{\alpha \beta C}{\bf T}^C
,\\
&&[{J}_{\alpha},{J}_{\beta}]_{{}_-}
={c}^{(1)}_{\alpha \beta \gamma}{J}^\gamma~;
\eea
\bea
&&[{J}_{\alpha},{\bf T}_{B}]_{+} =
d^{(1)}_{\alpha \gamma B} {J}^\gamma ~,\\
&&[J_{\alpha}, {\bf T}_{B}]_{-} =
c^{(2)}_{B D \alpha}{\bf T}^D~.
\eea

The coefficients $c^{(1)}_{\alpha\beta\gamma}$ are totally antisymmetric
with $c^{(1)}_{-+0}=1$.
The coefficients $d^{(1)}_{\alpha\beta,C}, d^{(2)}_{ABC}$ and
$c^{(2)}_{AB\gamma}$ are given in Tables 1 to 3.

\[
\begin{array}{|rrr|r|}\hline
 A & B & C & d^{(2)}_{ABC} \\ \hline\hline
 -2 &  2 &  0 & -1           \\
 -2 &  1 &  1 &  \sqrt{3/2}  \\
 -2 &  0 &  2 & -1           \\ \hline
 -1 &  2 & -1 &  \sqrt{3/2}  \\
 -1 &  1 &  0 & -1/2         \\
 -1 &  0 &  1 & -1/2         \\
 -1 & -1 &  2 &  \sqrt{3/2}  \\ \hline
  0 &  2 & -2 & -1           \\
  0 &  1 & -1 & -1/2         \\
  0 &  0 &  0 &  1           \\
  0 & -1 &  1 & -1/2         \\
  0 & -2 &  2 & -1           \\ \hline
  1 &  1 & -2 &  \sqrt{3/2}  \\
  1 &  0 & -1 & -1/2         \\
  1 & -1 &  0 & -1/2         \\
  1 & -2 &  1 &  \sqrt{3/2}  \\ \hline
  2 &  0 & -2 & -1           \\
  2 & -1 & -1 &  \sqrt{3/2}  \\
  2 & -2 &  0 & -1           \\ \hline
\end{array}
\quad
\begin{array}{|rrr|r|}\hline
 A & B & c & c^{(2)}_{ABc}\\ \hline\hline
 -2 &  2 &  0  & -2        \\
 -2 &  1 &  1  & -\sqrt{2} \\ \hline
 -1 &  2 & -1 & -\sqrt{2}  \\
 -1 &  1 &  0 &  1         \\
 -1 &  0 &  1 & -\sqrt{3}  \\ \hline
  0 &  1 & -1 & -\sqrt{3}  \\
  0 &  0 &  0 &  0         \\
  0 & -1 &  1 &  \sqrt{3}  \\ \hline
  1 &  0 & -1 &  \sqrt{3}  \\
  1 & -1 &  0 & -1         \\
  1 & -2 &  1 &  \sqrt{2}  \\ \hline
  2 & -1 & -1 &  \sqrt{2}  \\
  2 & -2 &  0 &  2         \\ \hline
\end{array}
\quad
\begin{array}{|rrr|r|}\hline
 a & b & C & d^{(1)}_{abC}  \\ \hline\hline
 -1 &  1 &  0 & -1/\sqrt{3}  \\
 -1 &  0 &  1 &  1           \\
 -1 & -1 &  2 & - \sqrt{2}   \\ \hline
  0 &  1 & -1 &  1           \\
  0 &  0 &  0 & -2/\sqrt{3}  \\
  0 & -1 &  1 &  1           \\ \hline
  1 &  1 & -2 & -\sqrt{2}    \\
  1 &  0 & -1 &  1           \\
  1 & -1 &  0 & -1/\sqrt{3}  \\ \hline
\end{array}
\]
Note that
\be
d^{(1)}_{abA} = \left({\bf T}_A\right)^{\alpha \beta}
e_{a}^{\alpha}e_{b}^{\alpha}~.
\ee
\subsection{Generators for the D-functions }

\noindent
The following five-dimensional spin matricies
has been used in the main text

\[
\left({\bf J}^{+}\right)_A^B =
\left (
\begin{array}{ccccc}
0    &  \sqrt{2}  &    0       &  0         &   0      \\
0    &   0        &  -\sqrt{3} &  0         &   0      \\
0    &   0        &    0       &  -\sqrt{3} &   0      \\
0    &   0        &    0       &  0         & \sqrt{2} \\
0    &   0        &    0       &  0         &   0
\end{array}
\right ),
\quad
\left({\bf J}^{-}\right)_A^B =
\left (
\begin{array}{ccccc}
0         &  0        &  0         & 0         &  0 \\
-\sqrt{2} &  0        &  0         & 0         &  0 \\
0         &  \sqrt{3} &  0         & 0         &  0 \\
0         &  0        &  \sqrt{3}  & 0         &  0 \\
0         &  0        &  0         & -\sqrt{2} &  0
\end{array}
\right ),
\quad
\left({\bf J}^{0}\right)_A^B =
\left (
\begin{array}{ccccc}
2  &  0  &  0  &  0  &  0 \\
0  &  1  &  0  &  0  &  0 \\
0  &  0  &  0  &  0  &  0 \\
0  &  0  &  0  & -1  &  0 \\
0  &  0  &  0  &  0  & -2
\end{array}
\right ).
\quad \quad
\]
The corresponding Cartesian components
$\left({\bf J}^{i}\right)_A^B:=
\eta^{\alpha\beta}e_\alpha^i\left({\bf J}_{\beta}\right)_A^B$
are
\[
\left({\bf J}^{1}\right)_A^B =
\left (
\begin{array}{ccccc}
 0 & -1            &  0           &  0           &  0 \\
-1 &  0            &  \sqrt{3/2}  &  0           &  0 \\
 0 &  \sqrt{3/2}   &  0           &  \sqrt{3/2}  &  0 \\
 0 &  0            &  \sqrt{3/2}  &  0           & -1 \\
 0 &  0            &  0           & -1           &  0
\end{array}
\right ),
\quad
\left({\bf J}^{2}\right)_A^B =
i\left(
\begin{array}{ccccc}
 0 & -1            &  0           &  0           &  0 \\
 1 &  0            &  \sqrt{3/2}  &  0           &  0 \\
 0 & -\sqrt{3/2}   &  0           &  \sqrt{3/2}  &  0 \\
 0 &  0            & -\sqrt{3/2}  &  0           & -1 \\
 0 &  0            &  0           &  1           &  0
\end{array}
\right ),
\quad
\left({\bf J}^{3}\right)_A^B =\left({\bf J}^{0}\right)_A^B
\]
satisfying the $ SO(3)$ algebra
\be
[{\bf J}_a,{\bf J}_b]=i\epsilon_{abc}{\bf J}_c~.
\ee
Note that
\be
c^{(2)}_{ABc}=i({\bf J}^c)_{AB}~.
\ee
We use the D-functions as representation of rotations in 3-space defined
in terms of Euler angles $\chi=(\theta,\psi,\phi)$
\be
D(\psi,\theta,\phi)= e^{-i\psi{\bf J}_3} e^{-i\theta{\bf J}_1}
e^{-i\phi{\bf J}_3}~.
\ee
They can be obtained from the corresponding 3-dimensional representation
(\cite{Brink}) via the formula
\be
D(\chi)_{AB}={1\over 2}\mbox{Tr}\left(R(\chi){\bf T}_A
R^T(\chi){\bf T}_B\right)~.
\ee

\subsection{Basis for symmetric matrices}

We use the orthogonal basis
$\alpha_A = ( \overline{\alpha}_i , \ \alpha^i )$
for symmetric matrices. They read explicitly
\[
\overline{\alpha}_1 = \left (
\begin{array}{ccc}
1      &   0      &    0     \\
0      &   0      &    0     \\
0      &   0      &    0
\end{array}
\right )~,
\quad \quad
\overline{\alpha}_2 = \left (
\begin{array}{ccc}
0      &   0      &    0     \\
0      &   1      &    0     \\
0      &   0      &    0
\end{array}
\right )~,
\quad \quad
\overline{\alpha}_3 = \left (
\begin{array}{ccc}
0      &   0      &    0     \\
0      &   0       &    0     \\
0      &   0      &    1
\end{array}
\right )~,
\]

\[
\alpha^{1} = \left (
\begin{array}{ccc}
0      &   0      &    0     \\
0      &   0      &    1     \\
0      &   1      &   0
\end{array}
\right )~,
\quad \quad
\alpha^{2} = \left (
\begin{array}{ccc}
0      &   0      &    1     \\
0      &   0      &    0     \\
1      &   0      &   0
\end{array}
\right )~,
\quad \quad
\alpha^{3} = \left (
\begin{array}{ccc}
0      &   1      &    0     \\
1      &   0      &    0     \\
0      &   0      &   0
\end{array}
\right )~.
\]
They obey the following orthonormality relations:
\be
\mbox{tr}\ (\overline{\alpha}_i \overline{\alpha}_j) = \delta_{ij}~,
\quad \quad
\mbox{tr}\ (\alpha_i \alpha_j) = 2 \delta_{ij}~,
\quad \quad
\mbox{tr}\ (\overline{\alpha}_i \alpha_j) = 0~.
\ee


\end{document}